\documentclass[aps,prl, twocolumn,preprintnumbers,amsmath,amssymb,superscriptaddress,nofootinbib]{revtex4-1}  
\usepackage[linktocpage,pagebackref=false,hidelinks]{hyperref}

\usepackage{amsmath,setspace,amsfonts,latexsym}
\usepackage{amssymb}
\usepackage{color}
\usepackage{epsfig}
\usepackage{graphicx}
\usepackage{slashed}
\usepackage{ulem}
\usepackage[export]{adjustbox}

\def\bea{\begin{eqnarray}}
\def\eea{\end{eqnarray}}
\def\beq{\begin{equation}}
\def\eeq{\end{equation}}

\begin{document} 

\title{Dark matter filtering-out effect\\ during a first-order phase transition}

\author{Dongjin Chway} 
\email{djchway@gmail.com}
\affiliation{Center for Theoretical Physics of the Universe, Institute for Basic Science (IBS), Daejeon, 34051, Korea}

\author{Tae Hyun Jung} 
\email{thjung0720@gmail.com}
\affiliation{Center for Theoretical Physics of the Universe, Institute for Basic Science (IBS), Daejeon, 34051, Korea}
\affiliation{Department of Physics, Florida State University, Tallahassee, FL 32306, USA}

\author{Chang Sub Shin} 
\email{csshin@ibs.re.kr}
\affiliation{Center for Theoretical Physics of the Universe, Institute for Basic Science (IBS), Daejeon, 34051, Korea}

\preprint{CTPU-19-37} 

\begin{abstract}
If the mass of dark matter is generated from a cosmological phase transition involving the nucleation of bubbles, the corresponding bubble walls can filter out dark matter particles during the phase transition.
Only particles with sufficient momentum to overcome their mass inside the bubbles can pass through the walls. 
As a result, the dark matter number density after the phase transition has a suppression factor $\exp(-M_\chi/2\tilde \gamma T)$,  where $M_\chi$ is the dark matter mass, and  $\tilde \gamma$ and $T$ are the Lorentz factor and  temperature of the incoming fluid in the bubble wall rest frame, respectively. 
Under certain assumptions, we show that the filtering-out process can naturally provide a large suppression consistent with the observed dark matter density for a wide range of dark matter masses up to the Planck scale. Since the first-order phase transition is the decisive ingredient in our mechanism, a new connection is made between heavy dark matter scenarios and  gravitational wave observations. 
\end{abstract}

\maketitle

\noindent {\bf Introduction}
Thermal freeze-out mechanism has been regarded as a standard way to explain the amount of dark matter (DM)\,\cite{Lee:1977ua}.
As the temperature of the Universe falls below the freeze-out temperature, DM is no longer in chemical equilibrium and its comoving number density is frozen to the value proportional to the inverse of the DM annihilation cross section. For the observed DM density, DMs need a sizable annihilation rate, roughly as large as the electroweak interaction rate.
We call such hypothetical DM particles as weakly interacting massive particles (WIMP).

Motivated by the WIMP paradigm, there have been lots of experimental studies to reveal the particle nature of DM.
Especially, direct detection experiments to observe scattering events between DMs and nucleons have enormously increased their sensitivities for the last decades. However, we have not yet obtained a convincing signal.
The absence of a direct detection signal provides strong constraints on the simple WIMP DM models with masses from GeV to TeV scale\,\cite{Aprile:2018dbl}.

Even if we can take refuge in heavy WIMP scenarios, there is a strong upper bound on DM mass within the freeze-out mechanism. 
The upper bound comes from that as the DM mass increases, the maximum value of the annihilation cross section allowed by the perturbative unitarity decreases and eventually gets smaller than the required value for the correct DM density. 
The unitarity bound implies the WIMP mass to be less than around 100 TeV\,\cite{Griest:1989wd, Smirnov:2019ngs}.

Therefore if the DM mass is heavier than 100 TeV, there should be an additional process to fit in the correct relic density. Along this direction, 
the pioneering works\,\cite{{Chung:1998ua,Chung:1999ve,Fedderke:2014ura,Chung:1998zb,Kuzmin:1998kk}} 
studied the role of the early matter domination and inflation periods to obtain the correct heavy DM relics.  
The freeze-in thermal production\,\cite{Kolb:2017jvz}, and the series of co-scattering processes\,\cite{Kim:2019udq} are also used to make the thermal heavy WIMP DM scenario viable.
One of the easiest ways to overcome the problem is the entropy injection to the SM after freezing out DM, which could originate from a supercooled phase transition of the Universe\,\cite{Hui:1998dc,Hambye:2018qjv, Baratella:2018pxi}.

Our scenario is based on the cosmological environment that is essentially the same as in Refs.\,\cite{Hui:1998dc,Hambye:2018qjv, Baratella:2018pxi}, i.e. DM acquires mass during a phase transition.
In this letter, we highlight the consequence of first-order phase transition followed by bubble dynamics and find a new application of the setup: {\it DM filtering-out effect}.
We also show that the parameter space of our scenario is fully separated from that of Refs.\,\cite{Hui:1998dc,Hambye:2018qjv, Baratella:2018pxi}.

\noindent{\bf Filtering Effect}
It is plausible that the DM mass is not a constant, but is dynamically generated from spontaneous breaking of symmetry such as Higgs mechanism or chiral symmetry breaking by strong dynamics. 
Then, when the Universe is hot enough, thermal effects restore the symmetry prohibiting the DM mass. As the temperature drops below the critical temperature, phase transition begins and DM gains nonzero mass. 

If the corresponding phase transition is first order, bubbles of the broken phase nucleate and expand during the phase transition. Since the symmetry is unbroken outside bubbles, the DM is still massless there, while inside the bubbles the symmetry is broken, resulting in a nonzero mass of DM. 

The mass gap between outside and inside the bubble is the key factor of filtering-out mechanism;
if the energy of a massless DM particle outside the bubble is smaller than the mass gap, it cannot enter the bubble because of the energy conservation.
DM particles that do not have enough energy are {\it filtered out}.

To be more quantitative, let us consider the wall rest frame where the particle flux is coming from the unbroken phase. The number of particles penetrating the bubble wall per area $\Delta A$ and a time interval $\Delta t$ is estimated by
\beq
\hspace*{-0.2cm}
\frac{\Delta N_{\rm in}}{\Delta A} =  \frac{g_\chi}{(2\pi)^3} \int d^3\vec p \int_{r_0}^{r_0 - \frac{p_r \Delta t}{|\vec p|}  } dr \, {\cal T}(\vec p)\,  \Theta (-p_r) f(\vec p ; \vec{x}),
\label{Nin}
\eeq
where $r_0$ denotes the bubble radius, ${\cal T}(\vec p)$ is the transmission rate,
$f(\vec p ; \vec{x})$ is a distribution function of particles at position $\vec x$ and momentum $\vec p$, 
$g_\chi$ is the DM particle degrees of freedom, and $\Theta(x)$ is the unit step function.
For simplicity, we take classical transmission rate ${\cal T}(\vec p) \simeq  \Theta (-p_z- M_\chi)$ where $M_\chi$ is the DM mass inside the bubble.

\begin{figure}[t] 
\begin{center}
\includegraphics[width=0.43\textwidth]{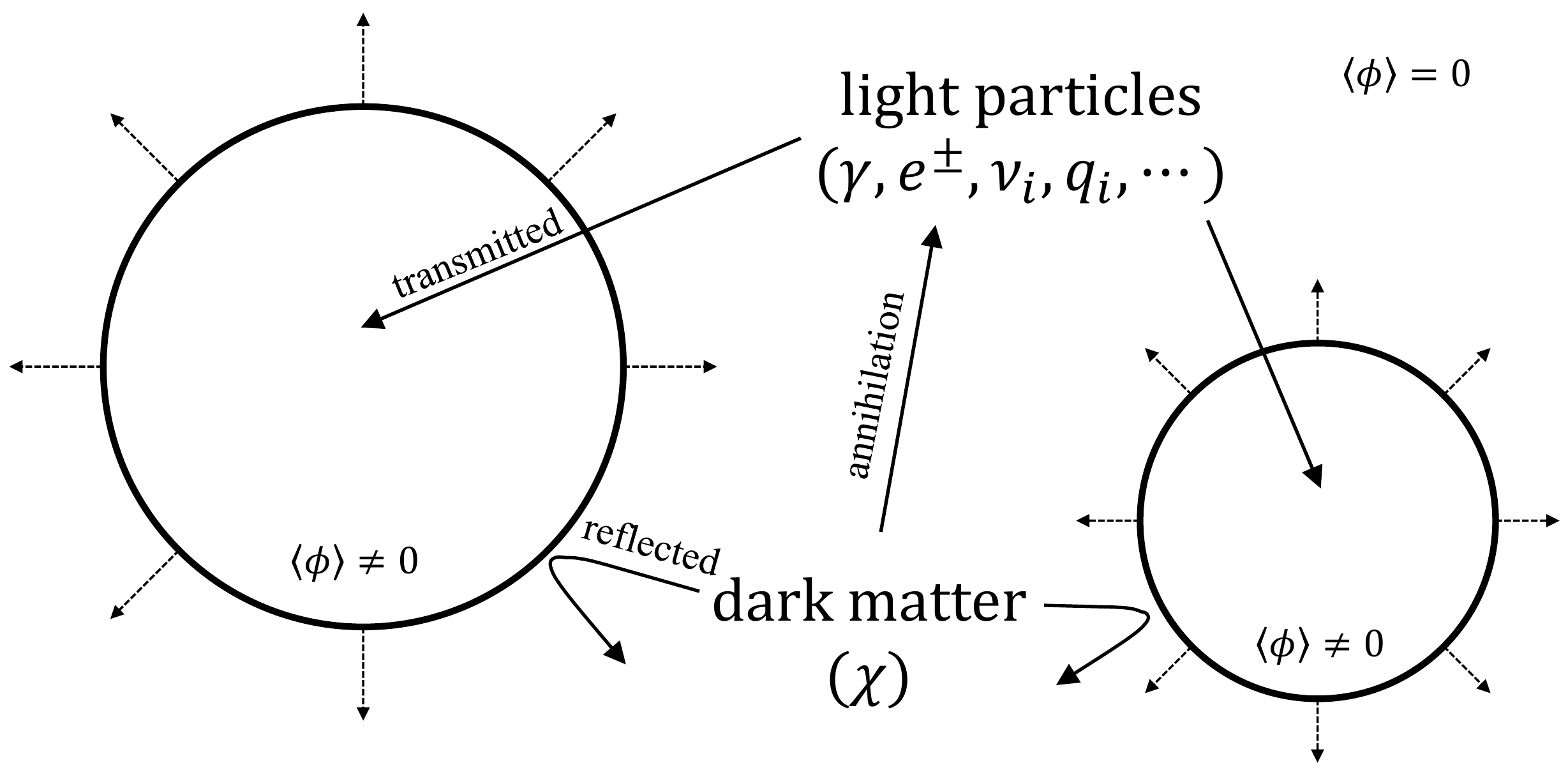} 
\end{center}
\caption{A schematic of the filtering-out mechanism. 
Most of DM particles ($\chi$) cannot penetrate through the bubble wall if momenta of particles outside the bubbles are not high enough to overcome the DM mass inside the bubbles. Outside the bubble, 
DM particles are massless and DM pair creation and annihilation processes are in thermal equilibrium. The light particles ($\gamma,\, e^\pm, \nu_i,\, q_i,\cdots$) which do not get sizable masses from bubbles can freely enter the bubbles.}
\label{fig:scheme}
\end{figure}

It is natural to assume that DM particles outside the bubble are thermalized even after scattering off the bubble walls since they are massless and have very short mean free-path and relaxation time.
In this case, $f$ follows the Bose-Einstein (BE) or the Fermi-Dirac (FD) distribution
\bea
f(\vec p;\vec x)\simeq f_{\rm eq}(\vec p, \vec {\tilde v},T)=\frac{1}{e^{\tilde \gamma (E-\vec {\tilde v} \cdot \vec p)/T}\pm1}.
\label{perfectfluid}
\eea
with $\tilde \gamma=1/\sqrt{1-\tilde v^2}$ being the Lorentz factor of the relative velocity $\tilde v$ of fluid bulk motion with respect to the wall.
For simplicity, we approximate it by the Boltzmann distribution, and obtain the flux $J^{\rm wall} =dN/(dA dt)$ as
\bea
 J^{\rm wall}&\simeq&  - g_\chi T^3 \left(\frac{ \tilde \gamma(1-\tilde v) M_\chi/T+ 1}{4\pi^2\tilde \gamma^3(1-\tilde v)^2}\right)e^{-\frac{\tilde \gamma(1-\tilde v) M_\chi}{T}}. 
\label{Jwall}
\eea
We obtain only $\mathcal{O}(0.2)$ change between BE and FD distributions.

The outgoing flux from bubbles is negligible in our scenario.
Because of the large DM mass inside bubbles, the DM amount from the sub-thermal production\,\cite{Hambye:2018qjv} is negligibly small compared to that from the bubble wall penetration. 
The contribution from the penetrated DM particles is also small since they must take time to change their direction after entering the bubble while the bubble expands rapidly at a sizable wall velocity.

The average number density inside the bubble can be obtained by
\bea
(n_\chi)_{\rm in} &=& -  \frac{ J^{\rm wall} }{\gamma_w \xi_w},
\label{nchi}
\eea 
where $\xi_w$ is the bubble wall velocity and $\gamma_w$ is its Lorentz factor.
When $\tilde \gamma \gg M_\chi/T$ and $\xi_w \simeq \tilde v$, Eq.\,\eqref{nchi} approaches to $g_\chi T^3/\pi^2$ which is the equilibrium number density for Boltzmann distribution outside the bubble.
It means that the bubble wall does not filter out DM particles at all in this limit.
Note that even for $M_\chi\gg T$, the exponent in Eq.\,\eqref{Jwall} is sensitive to $\tilde v$. 
As $\tilde v\to 1$, $\tilde\gamma\gg 1$ and the exponent approaches to $-M_\chi/2\tilde\gamma T$. On the other hand, in the case of $\tilde v\to 0$, $\tilde\gamma\to 1$ and the exponent becomes $- M_\chi/T$.

Before evaluating $\tilde v$, let us discuss what happens to DM particles that do not penetrate the bubbles. If the filtering out effect is efficient, the reflected DM particles might be accumulated around the bubble wall as the bubble expands.
In the wall rest frame, this situation can be described by the equation
\bea
\dot n_\chi \simeq \tilde v c T n_\chi^{\rm eq} + \langle \sigma v \rangle_T ((n_\chi^{\rm eq})^2-n_\chi^2),
\label{acc}
\eea
where $n_\chi$ is the local DM number density in front of the wall, $\langle \sigma v \rangle_T$ is the thermally averaged annihilation cross section and $n_{\chi}^{\rm eq}$ is the equilibrium number density.
Here, $c$ is an order one coefficient such that $1/(c T)$ describes a thermal diffusion length scale after reflection.
Since DMs are massless outside the bubble, $\langle \sigma v \rangle_T \sim 1/T^2$ and $n_{\chi}^{\rm eq} \sim T^3$. One can check that 
 the acuumulation term and the annihilation term in Eq.\,\eqref{acc} make a balance, so only ${\cal O}(1)$ enhancement of local number density from $n_\chi^{\rm eq}$ is possible. 
Since it does not change our conclusion, we neglect the accumulation effect to simplify the discussion.

\noindent
{\bf Fluid velocity} 
In general, $\tilde v$ is not coincident with the bubble wall velocity $\xi_w$, because 
it also depends on the fluid dynamics.
As described in Ref.~\cite{Espinosa:2010hh}, one can classify fluid profiles near the bubble wall in three categories: deflagration, detonation and hybrid.

When $\xi_w$ is lower than the sound speed $c_s\simeq 1/\sqrt{3}$, fluid hit by the bubble wall forms a shock-wave shell in front of the bubble wall, which is called deflagration case.
In this case, there is a parallel motion of fluid outside the bubble, so $\tilde v < \xi_w$. 
If $\xi_w$ is larger than $c_s$, detonation occurs, i.e. instead of the shock-front shell, fluid affected by the bubble wall follows behind the bubble wall.
A proper linear combination of detonation and deflagration profiles can also be a solution to the fluid equation.
This linear combination is called a hybrid profile.

Following the notation of Ref.~\cite{Espinosa:2010hh}, we denote incoming and transmitted fluid velocities (temperatures) in the wall rest frame by $v_+$ ($T_+$) and $v_-$ ($T_-$), respectively. For $\tilde v$ and $T$, we set  
\bea
\tilde v = v_+~\,\text{and}\,~T=T_+,
\label{assumption}
\eea
based on a reasonable assumption that the scattering rate between the reflected DM and background radiations is more important than the DM-DM self scattering rate. 
As a result, the incoming DM particles which {\it will} be filtered out are not much affected by the outgoing reflected DM particles which {\it were} already filtered out.

The parameters $\xi_w$, $\tilde v$, and $T$ can be evaluated from the equilibrium condition between pressures: $\Delta V=P$ where $\Delta V$ is the potential energy difference between false and true vacua at zero temperature and $P$ is the pressure on the wall.
In our scenario, the filtering-out process itself provides a dominant contribution to the pressure;
if a particle gains its mass $M_i$ inside the bubble and $ \tilde \gamma T \lesssim 0.2 M_i$, most of them in the incoming flux are reflected at the wall and exert pressure. 
The pressure can be represented as\footnote{
On the other hand, the pressure for $\tilde \gamma T \gtrsim M_\chi$ 
is discussed in detail in \cite{Bodeker:2009qy,Bodeker:2017cim}.}
\bea
P =  \frac{d}{3}(1+\tilde v)^3 \tilde  \gamma^2 T^4,
\label{equilibrium}
\eea
defining the effective degrees of freedom by
\bea
d \equiv   \frac{\pi^2}{30}\sum_{ 0.2 M_i > \tilde \gamma T }\left(N_i(B)+\frac{7}{8} N_i(F) \right),
\label{d}
\eea
where $N_i(B)$ ($N_i(F)$) stands for the number of bosons (fermions). 
This is nonzero in our scenario because of DM.

As a final remark of the section, we introduce two dimensionless parameters to simplify discussions as 
\bea
\alpha_n = \frac{\Delta V}{ \rho_{\rm rad}(T_n)}, \quad \lambda_{\rm eff}=\frac{\Delta V}{M_\chi^4},
\eea
where $T_n$ is the bubble nucleation temperature and 
\bea
\rho_{\rm rad}(T)=\frac{\pi^2}{30}\sum_{M_i < T}\left(N_i(B)+\frac{7}{8} N_{i}(F)\right) T^4. 
\label{a}
\eea 
The $\alpha_n$ represents how much the phase transition is being supercooled, and $\lambda_{\rm eff}$ shows how small $\Delta V$ should be in the unit of DM mass.

Using these variables, we can argue that it is difficult to realize the DM filtering-out mechanism for a small $\tilde v$, i.e. for $\tilde\gamma={\cal O}(1)$. 
Because of the condition $\Delta V=P$ and Eq.\,\eqref{equilibrium}, $\tilde\gamma={\cal O}(1)$ implies $\alpha_n \lesssim {\cal O}(1)$. 
At the same time, for a successful filtering-out process, we need a large $M_\chi/T_n\simeq$$2.4(\alpha_n/\lambda_{\rm eff})^{1/4}\sim{\cal O}(10)$ from Eq.\,\eqref{Jwall}. Thus, a very small $\lambda_{\rm eff}\lesssim{\cal O}(10^{-4})$ is required. However, in most cases, small $\lambda_{\rm eff}$ gives a suppression of nucleation rates, so the phase transition is quite delayed. This leads to a large $\alpha_n$ which is inconsistent with $\alpha_n\lesssim {\cal O}(1)$. 
For this reason, we will focus on large $\tilde v$ ($\tilde\gamma\gg 1$) with a large $\alpha_n$ cases in the rest of this letter.
\\

\noindent {\bf Bubble collsions}
During the bubble expansion, the potential energy stored in the symmetric vacuum is converted mostly into the bulk kinetic energy of the plasma fluid surrounding the bubble wall.
It is because the fluid pressure is equilibrated with the potential difference\,\cite{Espinosa:2010hh}. When bubbles collide, most of the bulk kinetic energy is converted into the thermal energy.\footnote{ The bubble collision also includes the collision of scalar field profiles which generates scalar field oscillations. When $\Delta V$ is large enough, the amount of heavy particles produced during the field collision is small~\cite{Falkowski:2012fb}.}

Let us consider what happens during bubble collisions.
Unlike scalar waves, fluid profiles cannot pass through each other since their mean free-path $\ell_{\rm fr} \sim \gamma/T$ is much smaller than the thickness of the plasma profile approximately given by the bubble size divided by $\alpha_n$.  
Lots of scattering events occur locally, resulting in a formation of hot plasma within the fluid collision surface.
This thermal energy will eventually be spread out over the space until the temperature becomes spatially homogeneous at the reheating temperature $T_{\rm rh}\simeq \alpha_n^{1/4} T_n$.  

The plasma bulk energy density in the bubble center rest frame is given by $\gamma_{\rm pl}^2 T_{\rm fluid}^4$ where $\gamma_{\rm pl}$ is the Lorentz factor of the fluid, and $T_{\rm fluid}$ is the temperature of the fluid profile. 
We have $\gamma_{\rm pl} \sim \gamma_w \propto \alpha_n^{1/2}$ from $\Delta V= P$, and $T_{\rm fluid}\sim \alpha_n^{1/4} T_n$\,\cite{Espinosa:2010hh}.
The maximum value of the local temperature at the moment of fluid collision is approximated by
\bea
T_{\rm max}\simeq \sqrt{2} \alpha_n^{1/2}T_n,
\eea
which is much higher than the reheating temperature $T_{\rm rh}\simeq \alpha_n^{1/4} T_n$ for large $\alpha_n$.  
In order to prevent additional DM production from the plasma collisions, we restrict our scenario to satisfy $T_{\rm max}<T_{\rm fo}$ where $T_{\rm fo}$ is the DM freeze-out temperature within the collision surface.

For the detonation fluid profile, the fluid collision happens inside the broken phase, and $T_{\rm fo}={\cal O}(0.1)M_\chi$. 
After collisions, DMs are chemically out-of-equilibrium for all time in the broken phase since $T_n< T_{\rm rh} < T_{\rm max} < T_{\rm fo}$.

On the other hand, for the hybrid profile, the bulk energy is concentrated on a shock wave in front of the bubble wall, so the fluid collision occurs outside bubbles.
Because DM is massless outside bubbles, $T_{\rm max}$ is always greater than $T_{\rm fo}$. 
This leads to a sizable production of high energetic DMs that can enter the broken phase without a large suppression factor. Thus, we exclude the hybrid fluid profile from our consideration. 
In a nutshell, the detonation fluid profile with a large $\alpha_n$ is the only possibility for the filtering-out scenario.
\\

\noindent {\bf Results}
Among the key parameters $\{M_\chi, d, \alpha_n, \lambda_{\rm eff}, T_n\}$ discussed so far, $M_\chi$ can be rewritten as $M_\chi \simeq 2.4(\alpha_n/\lambda_{\rm eff})^{1/4} T_n$. 
Therefore the DM relic density can be expressed by four parameters: $d$, $\alpha_n$, $\lambda_{\rm eff}$, and $T_n$. 
In the following estimation, we consider a fermionic DM with $g_\chi=2$, which is the main source of the pressure, so $d=2(7/8)(\pi^2/30)$.
We also fix $\rho_{\rm rad}(T)/T^4=106.75(\pi^2/30)$ which corresponds to the SM value at high temperature.

\begin{figure}[t] 
\begin{center}
\includegraphics[width=0.44\textwidth]{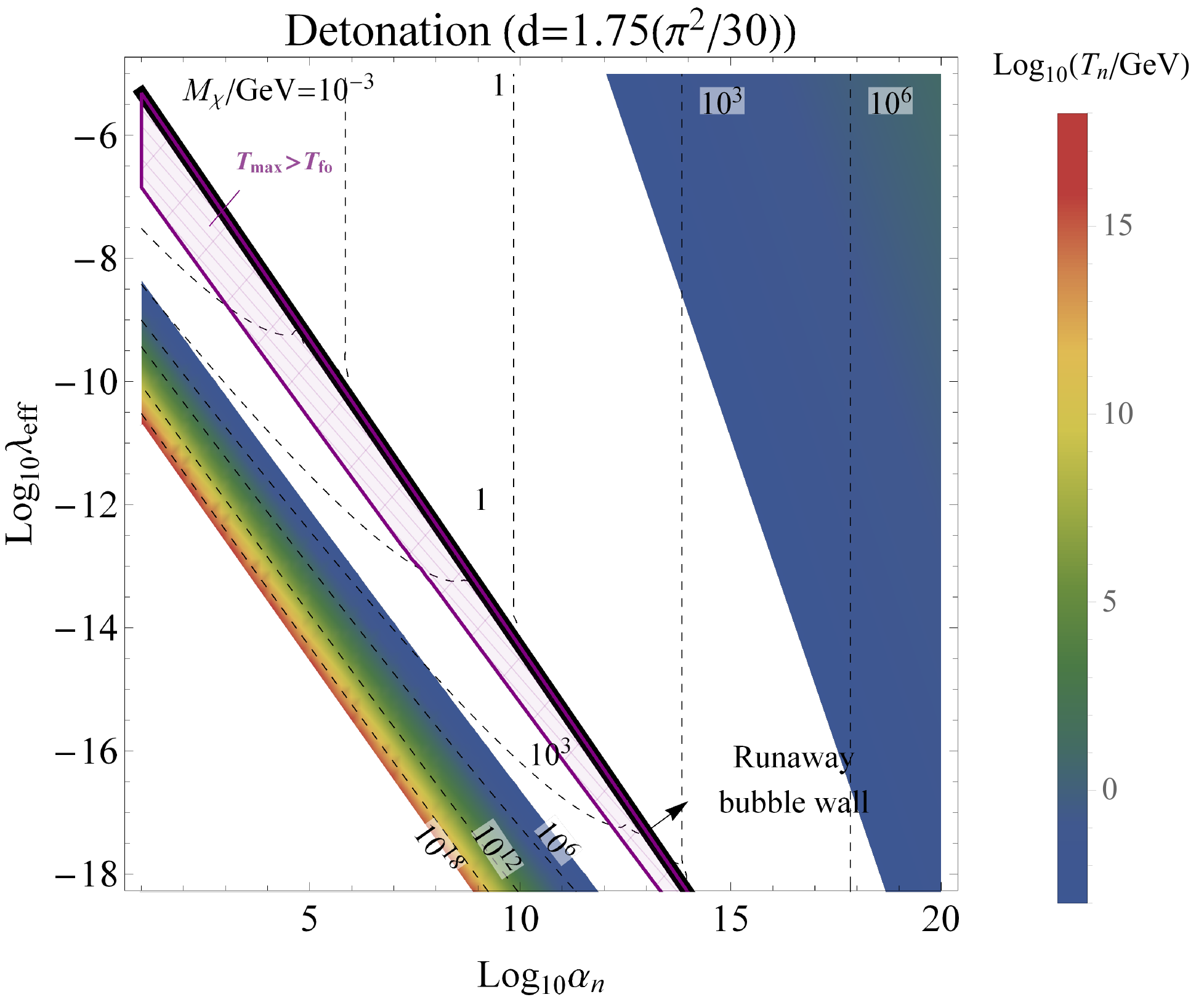} 
\end{center}
\caption{Required $T_n$ for the observed DM relic in the $\alpha_n$-$\lambda_{\rm eff}$ plane with the detonation profile. Dashed lines indicate the DM mass $M_\chi$.
The purple region is where the DM particles are produced too much during the bubble collisions.}
\label{fig:result1}
\end{figure}

With the detonation fluid profile ($T=T_n$, $\tilde \gamma=\gamma_w$), we show numerical values of $T_n$ required for the correct DM relic density in Fig.\,\ref{fig:result1}.
Only phenomenologically viable $T_n$ is marked by a color function, and each color represents $T_n$ whose value can be read from the bar legend on the right. 
The upper bound of $T_n$ is coming from the validity of effective theory, $T_n < T_{\rm rh} \ll M_{\rm Pl}$. For the lower bound, the successful Big Bang Nucleosynthesis requires $5\,{\rm MeV} \lesssim T_{\rm rh}$. 
For simplicity, we take a rather conservative bound as $5\,{\rm MeV} \lesssim T_n$ 
keeping same number of light degrees of freedom for all temperatures.

It is noteworthy that, in the figure, there are two viable regions that are separated and can be described by different physics.
In the diagonal band located at the bottom-left region, the filtering-out effect is the main part of determining the DM relic density.
Since $\tilde \gamma = \gamma_w\sim \alpha_n^{1/2}$ and $T=T_n$, the exponent in Eq.\,\eqref{Jwall} becomes $ -M_\chi/ \tilde \gamma \, T\sim -1/(\lambda_{\rm eff} \alpha_n )^{1/4}$.
This is why the figure shows a degenerate line along $\lambda_{\rm eff} \alpha_n={\rm const}$ and small change of $\lambda_{\rm eff} \alpha_n$ makes a big difference in $T_n$.
Corresponding DM mass can be read from the dashed lines and one can see that its values can be as large as the Planck scale in this region.

On the other hand, in the top-right corner, most of the DM particles just enter the bubble wall because the bubble wall runs away as indicated by the thick diagonal line in the middle.
In this case, the filtering-out effect does not affect DM relic density, but the dilution from the entropy production provides a large suppression $(T_n/T_{\rm rh})^3$ for the observed DM relic density as pointed out in Refs.\,\cite{Hui:1998dc,Hambye:2018qjv, Baratella:2018pxi}. 
The residual pair annihilation can further reduce the DM density, which is not discussed here because it is quite model dependent.
See also Ref.\,\cite{Hambye:2018qjv} for more refined estimation of DM relic in the cases with $\alpha_n \gtrsim 10^{20}$, $\lambda_{\rm eff}\gtrsim 10^{-5}$.

The purple shaded region describes where $T_{\rm max}>T_{\rm fo}$ so there can be a sizable DM production during bubble collisions.
One can see that the parameter space for the filtered-out DM scenario (rainbow-colored band in the bottom-left region) is safe from this criterion.
When bubble walls run away, 
the potential energy difference is mostly converted into the kinetic energy of scalar field
and those scalar waves passes through each other without generating local hot plasma at the collision surface\,\cite{Jinno:2019bxw}.
In conclusion, bubble collisions do not give a meaningful contribution to the DM abundance in both scenarios.

The previous result is based on the simple assumption Eq.\,\eqref{assumption}. This condition 
might not be valid if the DM self scattering rate is big enough.
In such a case, we can consider other possibilities under the condition that 
there is a terminal velocity of the bubble wall: 1) $\tilde v=v_-$ and $T=T_-$, and 2) $\tilde v=0$ and $T\simeq T_{\rm rh}$.
In both cases it turns out that the suppression factor $\exp[-\lambda_{\rm eff}^{-1/4}]$ 
does not depend on $\alpha_n$ and the dilution effect is absent since $T\sim T_{\rm rh}$. 
Here, we emphasize that in order to figure out correct boundary conditions near the bubble wall, it is important to solve full Boltzmann equations crossing the wall especially for the case with a strong DM self-interaction.\\

\noindent {\bf Discussion}
In this letter, we have investigated the possibility that the DM relic abundance is determined by the filtering-out effect of the bubble wall during a first-order phase transition.
We have shown that the DM number density after phase transition is suppressed by $\exp(-M_\chi/2\tilde \gamma T)$.
Unlike the freeze-out mechanism, 
our scenario does not have any theoretical lower bound of DM number density so that the DM mass can be as large as the Planck scale.
In terms of effective parameters $\alpha_n$ and $\lambda_{\rm eff}$, we find that $\lambda_{\rm eff} \alpha_n={\cal O}(10^{-9})$ is a good benchmark for the observed DM relic density.

An intrinsic observable of this mechanism is a gravitational wave signature since a strong first-order phase transition is required, $T_n\ll M_\chi$. 
The gravitational wave produced in a first-order phase transition has been widely studied in various contexts\,\cite{Kosowsky:1991ua, Kosowsky:1992rz, Kosowsky:1992vn, Kamionkowski:1993fg,Cutting:2018tjt, Ellis:2019oqb, Caprini:2019egz, Audley:2017drz}.
The signal peak frequency is, roughly, $1/\bar R$ multiplied by a redshift factor where $\bar R$ is the bubble radius at the bubble collision.
To estimate the signal strength, we need to specify a model, but it can be arbitrary at this moment.
If we have more observational information from future gravitational wave detectors\,\cite{Audley:2017drz, Graham:2016plp, Graham:2017pmn, Punturo:2010zz, Hild:2010id, Kawamura:2006up, Yagi:2011wg}, we will be able to narrow down $T_n$, $M_\chi$ and $\Delta V$ required.

As a final remark in the model building aspect, we note that the scalar potential should contain at least two different mass scales. 
Let us first consider 
a Mexican hat potential $V= -m^2 \Phi^2+\lambda \Phi^4$ which has only one massive parameter $m$.
Given a Yukawa coupling $y_\chi$ between $\Phi$ and the DM $(M_\chi = y_\chi \langle \Phi\rangle)$,
we have $\lambda_{\rm eff}\sim \lambda/y_\chi^4$, $T_n \sim m/y_\chi$
 and $\alpha_n\simeq 10^{-4}y_\chi^4 /\lambda$.
It results in $\lambda_{\rm eff}\alpha_n \simeq 10^{-4}$ which is too big compared to the benchmark value for the observed DM relic, ${\cal O}(10^{-9})$.

One of the working examples to provide multiple scales is the supersymmetric (SUSY) axion model in gauge mediation with a messenger scale $M \ll M_{\rm Pl}$. 
The shapes of the scalar potential for the saxion (the superpartner of the axion) field are quite different between two regions $\Phi < M$ and $\Phi > M$.
In the field range $\Phi < M$, soft SUSY breaking mass terms are generated by gauge mediation so that $V  \sim - m_s^2 \Phi^2$, while for $\Phi > M$ its effect is quite suppressed and the potential becomes $-m_s^2 M^2 (\ln \Phi/M)^n  + m_{3/2}^2 \Phi^2$ \cite{Moroi:2013tea,Asaka:1998ns,Asaka:1998xa,ArkaniHamed:1998kj,Abe:2001cg,Nakamura:2008ey, Jeong:2011xu,Choi:2011rs,Nakayama:2012zc}. Here, 
$n$ is the number of loops through which the mediator affects the potential, and $m_{3/2}$ is the gravitino mass that is much smaller than $m_s$, so the vacuum value of the saxion is evaluated as $\langle \Phi\rangle \sim M m_s/m_{3/2} \gg M$.  
In that case, $T_n\simeq m_s$, $\Delta V \sim m_s^2 M^2$, and 
one can easily obtain $  \lambda_{\rm eff} \alpha_n\sim  (m_{3/2}/m_s)^4={\cal O}(10^{-9})$.
We leave detailed studies within specific models as future works.
\\

\noindent {\bf Acknowledgement}
This work was supported by IBS under the project code, IBS-R018-D1. T.H.J. is supported by the US Department of Energy grant DE-SC0010102 and Prof. Kohsaku Tobioka's startup fund at Florida State University (Project id: 084011- 550- 042584).
\\
\\
\noindent {\bf Note added} 
While this paper was under completion, the Ref.\,\cite{Baker} appeared on arXiv, which is based on similar but different implementations.
\\

\end{document}